\documentclass[12pt, a4paper]{article}
\usepackage[english]{babel}
\usepackage[utf8]{inputenc}
\usepackage[autostyle, english = american]{csquotes}
\usepackage[T1]{fontenc}
\usepackage{times}
\usepackage{amsmath,amssymb,amsfonts}
\usepackage{graphicx}
\usepackage[dvipnames]{xcolor}
\usepackage{caption,subcaption}
\usepackage[margin=2cm]{geometry}
\usepackage[doublespacing]{setspace}
\usepackage{hyperref}
\usepackage{textgreek}
\usepackage[utf8]{inputenc}
\usepackage{authblk}
\usepackage{cleveref}
\usepackage[bottom]{footmisc}
\title{\vspace{-15mm} Quantum fluctuations of a regular charged black hole in massive gravity  }
\author[1]{Amruta Desai \footnote{amrutadesai011@gmail.com}}
\author[1]{Shubham Sharma \footnote{shubham2019gb@gmail.com}} \affil[1]{Department of  Physics, National Institute of Technology, Srinagar, Jammu \& Kashmir}
\author[1]{Prince A. Ganai \footnote{princeganai@nitsri.net}}
\date{June 2022}
\begin{document}
\maketitle
\begin{abstract}
 We examine the thermodynamics of a regular charged black hole (RCB) added with corrections due to massive gravity and thermal fluctuations at quantum level. We then derive the expressions for all the relevant thermodynamic quantities such as entropy, Hawking's temperature, internal energy, Gibbs free energy, corrected to first-order. We also briefly discuss the stability of such a system with the help of quantities like specific heat and thermodynamic pressure, along with this we present their comparative study with the corresponding original values.       
\end{abstract} 
\pagebreak
\section{Introduction}
Since the past few decades, black holes have been studied immensely as a potential connection that would relate two very crucial theories of modern physics, namely quantum mechanics and general relativity. This connection is credited to the second law of thermodynamics and its possible violation. It was thought that objects disappearing in the black hole would mean loss of information because black holes are characterised only by their mass, charge and angular momentum and they carry no information about the structure of matter that fall into them. This paradox was resolved by assigning black holes with a maximum allowed entropy \cite{1,2,3}. And this is the edifice on which the aforementioned black hole connection is built, i.e., the holographic duality \cite{4,5}. Along with this, we have quantum fluctuations that play a vital role in physics near the Planck scale, in a sense that it leads to non-trivial corrections to all the quantities and relations of utmost significance. Holographic principle, which relates black holes's entropy to its area, is certainly not an exception to this. There are two key aspects, rather closely related, that needs to be considered while discussing about these quantum fluctuations and black holes together :
\begin{itemize}
    \item The Bekenstein-Hawking area relation \cite{6,7}, whose significance can be marked by the way it connects the spacetime with the thermodynamics via Jacobson formalism \cite{8}, plays a pivotal role when it comes to corrections due to quantum fluctuations. Intuitively, there will be corrections to the relation $S = A / 4$, where A is the area of black hole, and in turn to the holographic principle \cite{9,10} as we approach quantum scales. As we go to higher energy levels, these quantum fluctuations have been shown to produce corrections of logarithmic form using different approaches \cite{11,12,13,14,15,16,17,18}
    \item Secondly, as the size of the black hole gets smaller, increase in its temperature leads to thermal fluctuations \cite{19}, which can be viewed as perturbations to the equilibrium thermodynamics, that would play a major role to thus produce corrections to the entropy-area relation. This is also true for evaporating black holes towards the end. Interestingly, these thermal fluctuations were also shown to produce the aforementioned logarithmic corrections to the entropy of a black hole \cite{20}
\end{itemize}
Now it appears that the quantum fluctuations of the spacetime could be closely related to the thermal fluctuations of a black hole, in a sense that they might be considered as two sides of the same coin. Work in this direction has been reported in \cite{8}, it has been shown that one could study the corrections to the thermodynamics of black holes by taking into account the modified entropy-area relation  $S = S_o \ + \ \alpha \log A \ + \ \gamma_1 A^{-1} \ + \ \gamma_2 A^{-2} \ + \ \dots \delta \exp{(-S_o)}$ \cite{21} where $\alpha, \gamma_1, \gamma_2$ \ and $\delta$ are the coefficients depending on various black hole parameters \footnote{Note that the scale at which these fluctuations are to be considered is very important because at exceedingly high temperatures even the manifold description of spacetime fails to make any sense. }. \par The other key concept involved in the discussion here is "Massive gravity". Simply put, massive gravity is the idea of giving a small mass to the graviton in order to attempt to resolve some of the longstanding problems in physics and in doing so modifying general relativity (GR) at large distances. Infact, massive gravity theories have garnered enormous interest of theorists in the field due to its intriguing theoretical and phenomenological prospects. The search for a viable massive deformation to GR dates back to the efforts of Fierz and Pauli (FP) \cite{22} in 1939, where they considered a linearized spin-2 field
over a Minkowskian background and added a mass term to it, famously called the Fierz-Pauli mass term. Consequently it was found out that a generic Lorentz-invariant massive deformation of gravity propagates 6 degrees of freedom (DoF) around Minkowski space, at the quadratic level. However the sixth mode was found to be a ghost, a mode propagating with negative kinetic energy.  The resolution to this came in the form of the dRGT mechanism \cite{23,24} quite recently, which leads to a nonlinear theory that avoids the infamous Boulware-
Deser ghost simply by choosing wisely the appropriate coefficients for the nonlinear mass terms.
\par In this paper, we will study the effects of these fluctuations at high-energy scales on a RCB in massive gravity coupled with nonlinear electrodynamics by adding the correction term of exponential form. The reason  we divert our attention to exponential corrections is that they have been shown to be of immediate importance when a quantum black hole system is considered. On the other hand, there are instances where the logarithmic corrections don't even appear for large area values \cite{25}. This makes the exponential term much more interesting and reliable when it comes to very high energy scales. As mentioned earlier, the usual perturbative description of the black hole would collapse for an extremely small size black hole and we would have to then consider the non-perturbative corrections, which corresponds to the non-equillibrium corrections to the entropy of the black hole. Also it is well established fact that the non-perturbative corrections that we here focus onto can be expressed in terms of the original entropy in its exponential form \cite{21,26}, which simply confirms the use of exponential corrections in such studies.     Subsequently, it has been shown that these exponential form corrections to the entropy arises in nonperturbative supergravity computations \cite{21,26,27,28} \footnote{However no such corrections have been observed in Loop Quantum Gravity (LQG) calculations yet. }. This indicates that one cannot neglect corrections of the exponential form. \par The paper is organised as follows :  We begin by presenting the form of RCB under consideration with appropriate massive gravity and quantum corrections of exponential form. In the next section, we present the exponentially corrected thermodynamics by deriving corrected values of entropy, Hawking's temperature, internal energy, Gibbs free energy. Alongside this, we highlight the influence of such thermal and massive corrections, especially at small horizon radius values, through graphical analysis for all the above mentioned quantities. In section \ref{section : 4}, we mark the role of exponential corrections in the existence of a holographic dual. In section \ref{section : 5}, we discuss the stability of the black hole in detail through quantities like specific heat, thermodynamic pressure and trace of the Hessian matrix. In the final section, we summarize our results.   
\section{Regular charged black hole in massive gravity}
We consider the most general static and sperically symmetric metric,
\begin{equation} \label{eq : 1}
    ds^2 = -f(r)\ dt^2 + f(r)^{-1}\ dr^2 + r^2 \left(d\theta^2 + \sin{\theta}^2 d\phi ^2 \right) 
\end{equation} where the metric function \emph{f}(r) is expressed as \cite{29}, 
\begin{equation}
 f(r) = 1 - \frac{2m_o\ (r)}{r} + m^2 c_1 r + 2m^2 c_2   
\end{equation} where $m_o\ (r)$ = $\frac{\sigma(r)}{\sigma_{\infty}}M$ \  is a mass function related to the mass of the black hole $M$ through  a distribution function $\sigma (r)$ \cite{30} such that $\sigma(r) > 0$ and $\sigma' (r) > 0$ for $r \geq 0$. Also it satisfies the condition $\frac{\sigma (r)}{r} \rightarrow 0$ as $\emph{r} \rightarrow 0$ and $\sigma_{\infty}$ is the normalization factor. Here $r_{+}$ and $r_{-}$ are the usual inner and outer horizons respectively that obeys the relation $r_{\pm} = 2m_o(r_{\pm})$ and can be obtained by considering $f(r)|_{r = r_{+}} = 0$,\  \emph{q} = 2$Q$\ is a constant related to the electric charge of the black hole, \emph{m} is a mass parameter related to massive gravity and $c_1$, $c_2$ are positive constants. We consider exponential-type distribution with $\sigma (r) = \exp\left({\frac{-q^2}{2Mr}}\right)$ \cite{30} and normalization factor $\sigma (\infty)$ = 1 to get the final form of the  general metric function as,
\begin{equation} \label{eq : 3}
 f(r) = 1 - \frac{2M}{r} \ \exp\left({\frac{-q^2}{2Mr}}\right) + m^2 c_1 r + 2m^2 c_2.       
\end{equation}
The metric approaches to Reissner-Nordstr\"{o}m metric asymptotically
\begin{equation*}
    f(r) = 1 - \frac{2m(r)}{r} \approx 1- \frac{2M}{r} + \frac{q^2}{r^2}
\end{equation*}
Also note that the distribution function $\sigma(r) \rightarrow 0$ as $r \rightarrow 0.$
\section{Exponentially corrected thermodynamics}
In this section, we study the effects of non-perturbative exponential corrections to the thermodynamics of the RCB system in massive gravity \ref{eq : 1} and thus derive expression for related thermodynamics quantities. We begin with the expression for the entropy term  corrected to first order,
\begin{equation}
    S = S_o + \delta e^{-S_o}
\end{equation} where $\delta$ is a constant with dimension of length and $S_o = \pi r_{+}^2$ is the zeroth-order entropy term without any corrections. As mentioned earlier,  this exponential correction term would be irrelevant for black holes with larger areas but would produce significant quantum effects at shorter distances. This can be better illustrated through a comparison between entropy curves for its original and  exponentially corrected values. Figure \ref{fig:1} below shows the behaviour of entropy in terms of horizon radius $r_+$. The dotted line represents the original entropy with $\delta = 0$ and the solid curves represent the corrected entropy for different values of correction parameter $\delta$ and mass related to massive gravity is taken to be m = 0.2 \footnote{ Note that here onwards  we derive all the expressions using m = 0.2}. We observe that for negative values of $\delta$ we get negative entropy values, which implies instability of the system and is in violation of the second law of thermodynamics. We thus restrict ourselves to $\delta > 0$ values of the correction parameter. The behaviour of the corrected entropy at smaller values of $r_+$ indicates the impact of the quantum corrections through strong fluctuations at this scale for varied values of $\delta$. It can also be seen that for larger values of $r_+$, entropy is an increasing function of $r_+$, which is in accordance with the entropy-area relation and the corrections does not make any difference. This ultimately implies that the thermal inconsistencies does not hold any significance for large black holes.  
\begin{figure}
    \centering
    \includegraphics[width=10cm,height=7cm]{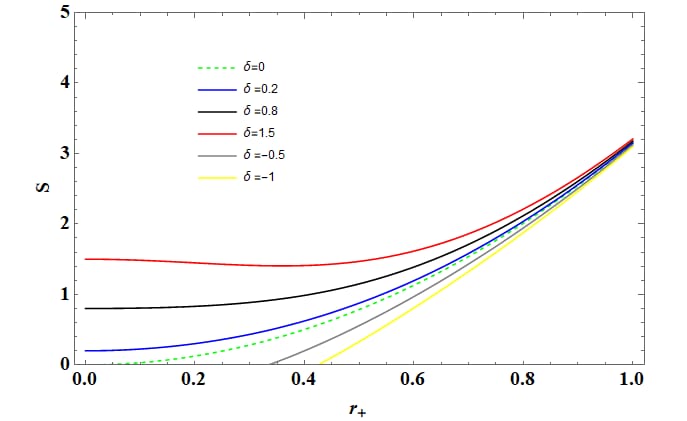}
    \caption{Entropy of a regular black hole in massive gravity in terms of horizon $r_+$ in comparison with its original value.}
    \label{fig:1}
\end{figure}

\begin{figure}
    \centering
    \includegraphics[width =10cm, height=7cm]{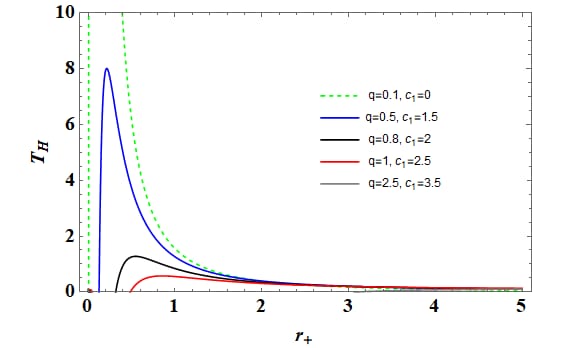}
    \caption{Hawking temperature for regular charged black hole in massive gravity in terms of horizon radius}
    \label{fig:2}
\end{figure}

\begin{figure} 
    \centering
    \includegraphics[width =10cm, height=7cm]{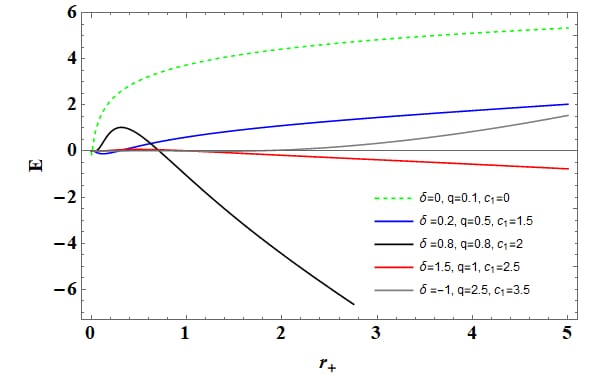}
    \caption{Internal energy function in terms of $r_+$}
    \label{fig: 3}
\end{figure}

The corresponding Hawking temperature is obtained by using \ref{eq : 3} and its relation with the metric function at $r_+$ as follows,
\begin{equation}
    T_H = \frac{1}{4 \pi} \left( \frac{df(r)}{dr}\right)_{r=r_+} =\frac{1}{4\pi}\Bigg[\frac{2M\exp\bigg(\frac{-q^2}{2Mr}\bigg)}{r^2}-\frac{q^2\exp\bigg(\frac{-q^2}{2Mr}\bigg)}{r^3} + c_{1}m^2\Bigg]
\end{equation}
Figure \ref{fig:2} illustrates the behaviour of Hawking temperature by variation of the electric charge and massive gravity parameter \emph{q} and \emph{c} respectively. For the case $q \neq 0 ,\ c \neq 0$, the temperature seems to vanish completely for some minimum value of radius (say $r_m$). This indicates that the black hole stops evaporating for $r_+ < r_m$, i.e., the Hawking radiations will not appear for values of $r_+$ below $r_m$. Thus we can say that information loss is not really a problem around this range \cite{31,32}. But at very small values of $r_+$, the temperature fluctuations appear, this can be attributed to the quantum fluctuations at this level which leads to the corresponding thermal fluctuations for very small black holes. This simply  confirms the relation between the quantum and thermal fluctuations at very high energy scales. 
Next, we find the internal energy function using the relation \footnote{Note that we have used $e^{-\pi r^2} \approx 1 - \pi r^2$ while deriving the expression.}, 
\begin{equation} \label{eq : 6}
E = \int T_H dS \ = \frac{(1 - \delta) \left[ -2M \exp{\left(\frac{-q^2}{2Mr}\right) + \frac{c_1}{2}m^2r^2 + 2\Gamma (0, \frac{q^2}{2Mr}})M\right]}{2}   
\end{equation}
The graphical analysis of internal energy function for different $r_+$ values (Figure \ref{fig: 3}) highlights the role of correction parameters in stabilizing our black hole system like no other quantity yet. It can be seen that for negative values of $\delta$, the internal energy is infact almost zero for small horizon values and further keeps on increasing with increasing size of the black hole. This positive value of internal energy indicates instability. On the other hand, for large $r_+$ values internal energy is decreasing function of all the correction parameters for $\delta > 0$. In contrary to the behaviour of the uncorrected internal energy, which is an increasing function of $r_+$, we notice that the black hole system under consideration would achieve more stability only when  the quantum  corrections are taken into account.  
We then move on to study the influence of the corrections on the   Helmholtz free energy function. It is calculated by using the expression for internal energy \ref{eq : 6}, 
\begin{equation}
    \begin{split}
\small F &= E - T_H S \\ &= \frac{(1 - \delta) \left[-2M \exp{\left(\frac{-q^2}{2Mr}\right) + \frac{c_1}{2}m^2r^2 + 2\Gamma (0, \frac{q^2}{2Mr}})M\right]}{2} \\ & \frac{1}{4\pi}\Bigg[\frac{2M\exp\bigg(\frac{-q^2}{2Mr}\bigg)}{r^2}-\frac{q^2\exp\bigg(\frac{-q^2}{2Mr}\bigg)}{r^3} + c_{1}m^2\Bigg] \left[ \pi r^2 + \delta e^{-\pi r^2 }\right].     
\end{split}
\end{equation}
\begin{figure}
    \centering
    \includegraphics[width =10cm, height=7cm]{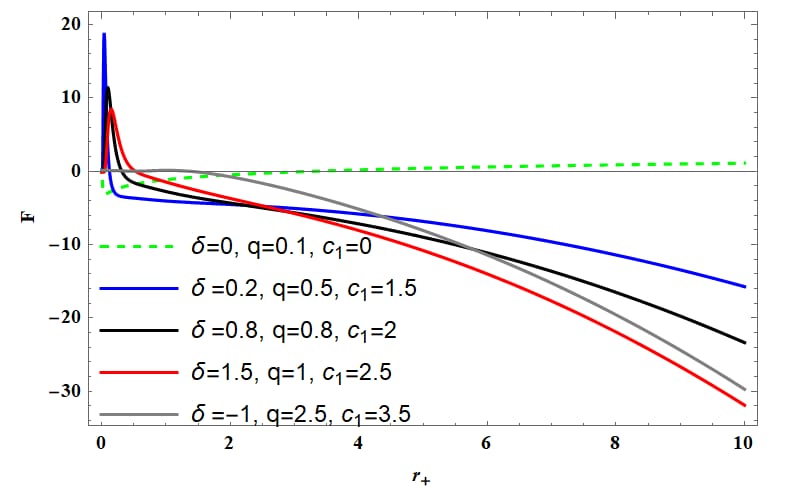}
    \caption{Helmholtz free energy in terms of $r_+$}
    \label{fig:4}
\end{figure} \\ Figure \ref{fig:4} above for the free energy function puts more light into the discussion on stability of the RCB simultaneously coupled with both electrodynamics and massive gravity. We observe that such a system is stable only for very small region of space with $r_+ < 1$. However when corrections are considered, the system can be stabilised for considerably larger region of space. For positive values of $\delta$ and other correction parameters, Helmholtz free energy decreases which points to stability of the system. But for negative value of $\delta$, Helmholtz free energy displays an anomalous behaviour, i.e., it has an increasing value for smaller $r_+$ region and then continues with the usual decreasing manner for $r_+ >2$. This could be attributed to the spacetime curvature manifesting itself in the form of tidal forces that subdues the outcome of corrections at quantum level. The same effect would be more apparent in the pressure case we discuss next. \\
Now we will derive the expression for modified thermodynamic pressure using following relation,
\begin{equation}
\begin{split}
P &= -\left(\frac{dF}{dV} \right)_{T_H} \\ &= - \frac{1}{4 \pi r^2} \\ \Bigg(&\frac{(1 - \delta) (\pi^2 + \delta e^{-\pi^2 })\Bigg[\frac{2M\exp\big(\frac{-q^2}{2Mr}\big)}{r^2}-\frac{q^2\exp\big(\frac{-q^2}{2Mr}\big)}{r^3} + c_{1}m^2\Bigg] \left[\frac{2M \exp\left(\frac{-q^2}{2Mr}\right)}{r^2}  - \frac{q^2 \exp\left(\frac{-q^2}{2Mr}\right)}{r^3} + c_1\ m^2 r \right]}{8 \pi} 
+ \\  &\frac{(1 - \delta)r \left[-2M \exp{\left(\frac{-q^2}{2Mr}\right) + \frac{c_1}{2}m^2r^2 + 2\Gamma (0, \frac{q^2}{2Mr}})M\right]\left[\frac{2M \exp\left(\frac{-q^2}{2Mr}\right)}{r^2}  - \frac{q^2 \exp\left(\frac{-q^2}{2Mr}\right)}{r^3} + c_1\ m^2 \right]  }{4} \ + \\ &\frac{(1 - \delta) (\pi r^2 + \delta e^{-\pi^2 })\left[-2M \exp{\left(\frac{-q^2}{2Mr}\right) + \frac{c_1}{2}m^2r^2 + 2\Gamma (0, \frac{q^2}{2Mr}})M\right]}{8 \pi} \times \\  &\left[\frac{-4M \exp\left(\frac{-q^2}{2Mr}\right)}{r^3}  - \frac{4q^2 \exp\left({\frac{-q^2}{2Mr}}\right)}{r^4} - \frac{q^4 \exp\left({\frac{-q^2}{2Mr}}\right)}{2Mr^5} \right]\Bigg).   
\end{split}    
\end{equation} From figure \ref{fig: 5} below, we can estimate the effect of massive gravity corrections on the pressure of the RCB. This quantity holds a special place in the discussion of thermodynamics of a black hole due to its vital connection with the spacetime geometry itself. For a black hole, the pressure is closely related to its cosmological constant which indicates that its the spacetime curvature value that would dominate over the thermal fluctuation or quantum correction effects, irrespective of the size of the black hole. We can see that for positive values of $\delta$, pressure decreases for small values of volume. But for $V < 1$ the pressure curves does not change with volume. This simply asserts our hypothesis that for the region $V > 1$ its the tidal forces associated with BH that control the behaviour of thermodynamic pressure and not any quantum corrections or thermal fluctuations. \par For $V >1$, the pressure values are increasingly positive, which implies stability. However for negative value of $\delta$, there exists a region of space $3 < V < 3.5$ where the pressure diverges and hence remains undefined. For $V > 3.5$, the pressure first exponentially increases but then asymptotically decreases, which again point towards instability. Note that this has been the case for all $\delta < 0$ values for entropy, internal energy and Helmholtz free energy.   
\begin{figure}
    \centering
    \includegraphics[width =10cm, height=7cm]{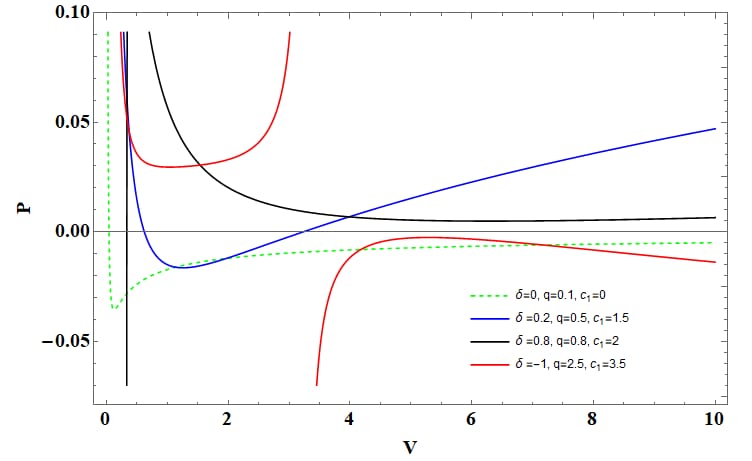}
    \caption{Pressure in terms of V}
    \label{fig: 5}
\end{figure} \\
Next we go on to calculate the corrected enthalpy of the BH system under consideration using the following relation,
\begin{equation}
\begin{split}
    H &= E + PV \\ &= \frac{(1 - \delta)\left[-2M \exp{\left(\frac{-q^2}{2Mr}\right) + \frac{c_1}{2}m^2r^2 + 2\Gamma (0, \frac{q^2}{2Mr}})M\right] }{2} - \\ & \frac{r(1 - \delta)(\pi r^2 + \delta e^{-\pi^2})\left( \frac{2M \exp{\left(\frac{-q^2}{2Mr}\right)}}{r^2} - \frac{q^2 \exp{\left(\frac{-q^2}{2Mr}\right)}}{r^3} + c_1 m^2\right)\left(\frac{2M \exp{\left(\frac{-q^2}{2Mr}\right)}}{r} - \frac{q^2 \exp{\left(\frac{-q^2}{2Mr}\right)}}{r^2} + c_1 m^2 r\right)}{24\pi} + \\ &\frac{r^2(1 - \delta) \left( -2M \exp{\left(\frac{-q^2}{2Mr}\right)} + \frac{c_1 m^2r^2}{2} + 2\Gamma\left(0,\frac{q^2}{2Mr}\right)M\right)\left( \frac{2M \exp{\left(\frac{-q^2}{2Mr}\right)}}{r^2} - \frac{q^2 \exp{\left(\frac{-q^2}{2Mr}\right)}}{r^3} + c_1 m^2 \right)}{12} + \\ &\frac{r(1 - \delta)(\pi r^2 + \delta e^{-\pi^2})\left(-2M \exp{\left(\frac{-q^2}{2Mr}\right)} + \frac{c_1 m^2r^2}{2} + 2\Gamma \left(0, \frac{q^2}{2Mr}\right)M \right)}{24\pi} \times \\ & \left( \frac{-4M \exp{\left(\frac{-q^2}{2Mr}\right)}}{r^3} + \frac{4q^2 \exp{\left(\frac{-q^2}{2Mr}\right)}}{r^4} - \frac{q^2 \exp{\left(\frac{-q^2}{2Mr}\right)}}{2Mr^5}\right).
\end{split}    
\end{equation}
The corrected Gibbs free energy can also be obtained using a similar relation,
\begin{equation}
    \begin{split}
G &= F + PV \\ &= \frac{(1 - \delta)\left[-2M \exp{\left(\frac{-q^2}{2Mr}\right) + \frac{c_1}{2}m^2r^2 + 2\Gamma (0, \frac{q^2}{2Mr}})M\right]}{2} - \\ &\frac{1}{4\pi}\left[\frac{2M\exp{\left(\frac{-q^2}{2Mr}\right)}}{r^2} - \frac{q^2 \exp{\left(\frac{-q^2}{2Mr}\right)}}{r^3} + c_1m^2 \right][\pi r^2 + \delta e^{-\pi r^2}] \ - \\ & \frac{r(1 - \delta)(\pi r^2 + \delta e^{-\pi^2})\left( \frac{2M \exp{\left(\frac{-q^2}{2Mr}\right)}}{r^2} - \frac{q^2 \exp{\left(\frac{-q^2}{2Mr}\right)}}{r^3} + c_1 m^2\right)\left(\frac{2M \exp{\left(\frac{-q^2}{2Mr}\right)}}{r} - \frac{q^2 \exp{\left(\frac{-q^2}{2Mr}\right)}}{r^2} + c_1 m^2 r\right)}{24\pi} + \\ &\frac{r^2(1 - \delta) \left( -2M \exp{\left(\frac{-q^2}{2Mr}\right)} + \frac{c_1 m^2r^2}{2} + 2\Gamma\left(0,\frac{q^2}{2Mr}\right)M\right)\left( \frac{2M \exp{\left(\frac{-q^2}{2Mr}\right)}}{r^2} - \frac{q^2 \exp{\left(\frac{-q^2}{2Mr}\right)}}{r^3} + c_1 m^2 \right)}{12} + \\ &\frac{r(1 - \delta)(\pi r^2 + \delta e^{-\pi^2})\left(-2M \exp{\left(\frac{-q^2}{2Mr}\right)} + \frac{c_1 m^2r^2}{2} + 2\Gamma \left(0, \frac{q^2}{2Mr}\right)M \right)}{24\pi} \times \\ & \left( \frac{-4M \exp{\left(\frac{-q^2}{2Mr}\right)}}{r^3} + \frac{4q^2 \exp{\left(\frac{-q^2}{2Mr}\right)}}{r^4} - \frac{q^2 \exp{\left(\frac{-q^2}{2Mr}\right)}}{2Mr^5}\right).
    \end{split}
\end{equation}

\section{Holography}
\label{section : 4}
\begin{figure}
    \centering
    \includegraphics[width =10cm, height=7cm]{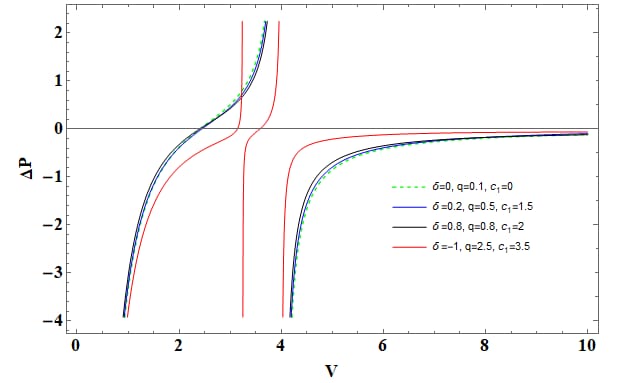}
    \caption{$\Delta P = P - P_w$ in terms of V}
    \label{fig: 6}
\end{figure}

It would be interesting to discover that the "holographic duality" exists for regular black holes in massive gravity. This idea, originating from the relation between black hole entropy and its area   \cite{6} has been applied to the gravitational spacetime as well \cite{33,34,35}. Subsequently, this common connection of holography between black hole physics and gravitational physics has often been employed to resolve some of the longstanding black hole mysteries. Thus, it would be of great consequence to observe that the RCB in massive gravity in fact has a holographic dual in the form of Van der Waal fluid that obeys the relation \cite{36} \footnote{Note that holographic duality has been observed in case of Born-Enfeld AdS black hole \cite{37}.}, 
\begin{equation} \label{eq : 9}
    \left(P_{\emph{w}} + \frac{a}{V^2}\right) \left( V - A_1\right) = T,
\end{equation} This is the equation of state (EoS) for a Van der Waal fluid with Boltzmann constant $K_B = 1$. Here \emph{a} is the intermolecular force strength and \emph{$A_1$} is the excluded volume. Equation \ \ref{eq : 9} can also be written for an ideal gas ( \emph{a} = \emph{$A_1$} = 0) \ as 
\begin{equation}
P_{\emph{w}} = \frac{T}{V - A_1} - \frac{a}{V^2}    
\end{equation} For the holographic duality to exist for an exponentially corrected BH in massive gravity, the black hole pressure must obey the relation $P = P_w$,\ i.e., $\Delta P = P - P_w = 0$. This can be clearly seen from the plot of $\Delta P$ in terms of \emph{V} \footnote{Note that we have considered a = $A_1$ = 4 while deriving the expression for $\Delta P$. }. Figure \ref{fig: 6} illustrates that $\Delta P = 0$ condition is satisfied for larger values of \emph{V} beyond 10. This indicates that thermal fluctuations at larger values of \emph{V} does not lead to breaking the holographic duality and hence does not influence the existence of a holographic dual for large black holes. The volume is diverging at several points which indicates at the possibility of Van der Waals dual in the presence of exponential corrections. 

\section{P - \emph{V} criticality and Black hole stability}
\label{section : 5}
\begin{figure}
    \centering
    \includegraphics[width =10cm, height=7cm]{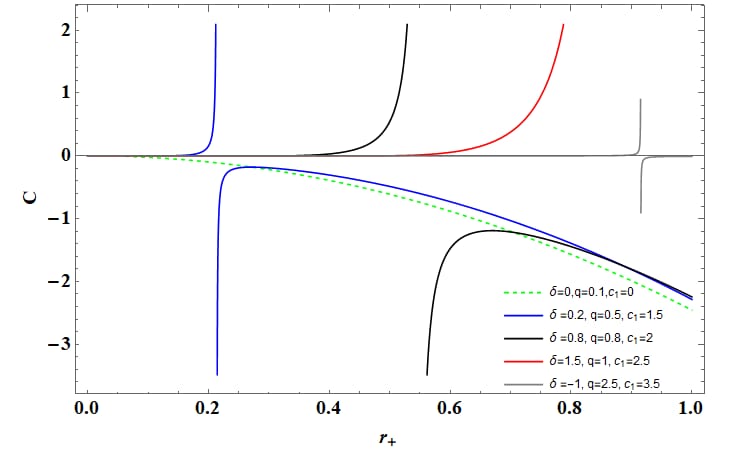}
    \caption{Specific heat in terms of $r_+$}
    \label{fig: 7}
\end{figure}

In order to fully understand the effect of thermal fluctuations on the stability of BH, we need to analyse its \emph{P} - \emph{V} behaviour in a different light. This is done using the relations,
\begin{equation}
    \begin{split}
    \left ( \frac{\partial P}{\partial V}\right)_{T = T_c} &= 0, \\ \left( 
    \frac{\partial^2P}{\partial V^2}\right)_{T = T_c} &= 0.
    \end{split}
\end{equation} The plot of \emph{P} in terms of \emph{V} (Figure \ref{fig: 5}) above suggests the presence of critical points for the RCB in massive gravity with exponential corrections \footnote{The \emph{P} - \emph{V} criticality for AdS black holes in massive gravity with logarithmic corrections are studied \cite{38}.}
As discussed earlier, thermal fluctuations play a vital role in the stability of the BH due to the existence of critical points only in the presence of exponential correction. To study the stability of the system further, we begin by deriving the expression for specific heat using the relation,
\begin{equation}
    C = T \left( \frac{dS}{dT}\right)
\end{equation}
which yields, 
\begin{equation}
\begin{split}
    C &= -\frac{r}{2} \Biggl[ \left( \frac{2M \exp{\left( \frac{-q^2}{2Mr}\right)}}{r^2} - \frac{q^2 \exp{\left( \frac{-q^2}{2Mr}\right)}}{r^3} + c_1m^2 \right ) (1 - \delta e^{-\pi r^2}) \ \times \\ &\left(\frac{8 \pi M r^5}{(8M^2r^2 - 8M q^2r + q^4) \exp{\left(\frac{-q^2}{2Mr}\right)}}\right)\Biggr]
\end{split}
\end{equation} \\ Figure \ref{fig: 7} above demonstrates the behaviour of specific heat with changing horizon radius $r_+$. For specific heat without any corrections, its values are all negative which signals BH instability. However when exponential corrections and massive gravity corrections are taken into account, specific heat has exponentially increasing behaviour starting from zero, i.e., positive values indicating stability. We also observe that the phase transition occurs only for the corrected values of specific heat, at a critical radius value (say \emph{$r_c$}). It is important to point out that positive specific heat values exists for small horizon radius, which implies stability for small black holes only possible due to the effects of correction.
Moreover it is important to consider a different approach to fully acknowledge the role exponential corrections in the stability of the model, i.e., by taking into account the Hessian matrix of the Helmholtz free energy. The expression for the same can be obtained by using the relation, \hspace{0.5cm}
\begin{equation} \label{eq : 16}
   \mathcal{H}  = 
\begin{pmatrix}
  \frac{\partial^2F}{\partial T^2} & \frac{\partial^2F}{\partial T \partial\mu} \\ \frac{\partial^2F}{\partial\mu \partial T} & \frac{\partial^2F}{\partial\mu^2}
\end{pmatrix}
\end{equation} For a positive eigenvalue of the Hessian matrix, the black hole is said to be thermodynamically unstable. Also, it is crucial to calculate the trace of the matrix \ref{eq : 16} in order to recognise the regions of black hole stability. This can be done by using the formula,
\begin{equation}
\tau = Tr (\mathcal{H}) = \left( \frac{\partial^2F}{\partial T^2}\right) + \left(\frac{\partial^2F}{\partial \mu^2}\right)    
\end{equation} The black hole that obeys the condition $\tau \geq 0$ is said to be stable. The expression for $\tau$ turns out to be,
\begin{equation}
\begin{split}
    \tau &= -\biggl[ \left(8\pi \delta c_1M^2m^2r^7e^{\frac{-q^2}{2Mr}} + (16\pi - 16\pi\delta)M^3r^5 + (24\pi\delta - 40\pi)M^2q^2r^4 + (14\pi - 4\pi\delta)Mq^4r^3 - \pi q^6r^2\right) \\ &\times e^{\pi r^2} + (16\pi^2 \delta c_1M^2m^2r^9 - 8\pi \delta c_1 M^2m^2r^7)e^{\frac{q^2}{2Mr}} + 32\pi^2\delta M^3r^7 - 16\pi^2\delta M^2 q^2r^6 + 48\pi\delta M^3r^5 - \\ &56\pi \delta M^2q^2r^4 + (8\pi\delta Mq^4 + 48 \delta M^3)r^3 - 72\delta M^2q^2r^2 + 18\delta Mq^4r - \delta q^6\biggl]e^{\left(-\pi r^2 - \frac{-q^2}{2Mr}\right)} \times \\ & \left[ \frac{16 \pi M^2r^7}{(48 M^3r^3 - 72 q^2r^2M^2+ 18M q^4r-q^6)e^{\frac{-q^2}{2Mr}}} - \frac{(2Mr - q^2)^3}{16 q M^3}\right]/ (16\pi M^2r^7)
\end{split}
\end{equation}
\begin{figure}[htp]
    \centering
    \includegraphics[width =10cm, height=7cm]{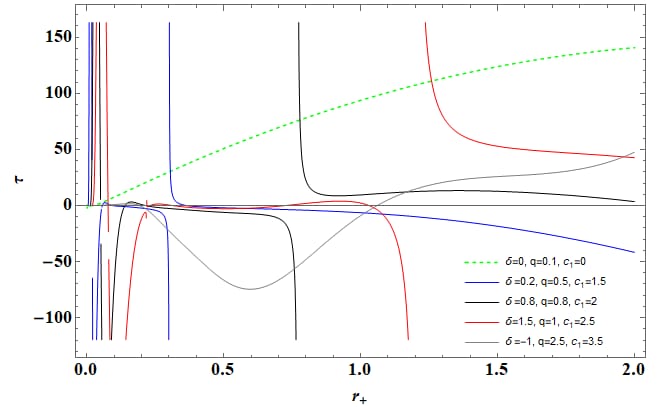}
    \caption{Trace of Hessian in terms of $r_+$}
    \label{fig: 8}
\end{figure} We make some further comments on the stability of the system by analysing plot \ref{fig: 8} of the trace of the Hessian matrix for varied $r_+$ values. Here we taken mass of the black hole $m_0 = 1$ to obtain the expression of $\tau$. For uncorrected case the system is already found to be stable for both smaller and larger BHs. And for the case with corrections, the system is stable for positive values of corrections parameter and unstable for negative value of $\delta$  at small $r_+$ values, which has been the case all along.

\pagebreak

\section{Conclusions}
In this paper, we have launched into a detailed investigation of the effects of non-perturbative corrections of exponential form on the RCB paired with massive gravity, at quantum level. The study of thermal fluctuations for such a black hole plays a key role here due to its connection with the quantum corrections, made apparent through the Jacobson formalism. We elaborate that this kind of first-order corrections to the thermodynamics of the black hole indeed produce some non-trivial modifications at the small horizon values. \par Firstly, we calculate the corrected entropy and the Hawking's temperature for such a system and analyse the effect of such quantum corrections through a  plot in terms of the horizon radius $r_+$. We find out that the corrected entropy is an increasing function of $r_+$, as expected and that there exists a region of space where the information paradox is actually resolved. However at very small $r_+$ values, small temperature fluctuations are visible. We then further calculate all the other relevant thermodynamic quantities such as the internal energy function, Helmholtz free energy, Gibbs free energy function and the corrected enthalpy. We observe an exponential growth in the internal energy near the small horizon values which can be attributed to the non-perturbative corrections at the quantum level. This unusual behaviour of the internal energy function clearly marks the significance of these corrections at very high energy scales. These corrections greatly influence the stability of the black hole at small horizon values as is evident from the behaviour of the Helmholtz free energy function. Finally, we also confirm the existence of a holographic dual in the form of Van der Waals fluid to this black hole in massive gravity. A detailed account of the stability of such a system is presented by looking into the presence of critical points through graphical analysis of $P-V$ behaviour of the black hole, towards the end of the paper.  

\end{document}